\documentstyle[11pt,paspconf,epsf]{article}

\begin{document}

\title{Photometric Redshifts: A Perspective from an Old-Timer\altaffilmark{1} on
	Its Past, Present, and Potential}

\author{David C. Koo}
\affil{UCO/Lick Observatory, Department of Astronomy and Astrophysics, University of California,
    Santa Cruz, CA 95064}


\altaffiltext{1}{Feedback from R. J. Weymann on receiving this title: ``Give me a break!!!''} 



\begin{abstract}

I review the early history of photometric redshifts; specify a working
definition that encompasses a broader range of approaches than
commonly adopted; discuss the pros and cons of template fitting versus
empirically-based techniques; and summarize some past
applications. Despite its relatively long history, the technique of
photometric redshifts remains far from being a mature tool. Areas
needing development include the use of spatial structure, the
incorporation of large redshift samples with multicolor photometry for
empirical calibrations of redshift errors, and improved analysis tools
that directly include redshift probability distributions rather than
singular values. Photometric redshifts has not only undergone a
recent revival -- it is also rapidly becoming a crucial tool of mainstream
observational cosmology.

\end{abstract}




\section{Brief Early History}

Photometric redshifts is a term that has received recent, wide-spread
visibility and interest, though the use of multicolor photometry to
estimate the redshifts of galaxies has a history that is relatively
long.  As of March 1999, my search of the ADS abstracts for the term
``photometric redshifts'' yielded only 9 papers ending in 1994, with
the first in 1982, i.e. averaging less than one paper per year. A
sudden jump then occurs in 1995 with 6 papers, followed by 1996 with
7, 1997 with 9, and then another major jump to 35 papers in 1998. The
broad participation and range of topics of this workshop validate this
surge in activity.

Among these papers, the earliest reference to the term ``photometric
redshifts'' is found in the abstract of Puschell et. al. (1982). Their
attempt to estimate redshifts of faint radio galaxies via broadband
photometry was pioneering (even by today's standards)  in three
respects: 1) the use of near-infrared bands ($JHK$) along with optical
bands ($RI$), 2) the adoption of $\chi^2$ fits to spectral energy
distributions (SED), and 3) the use of different sets of SED
templates, ranging from non-evolving local ellipticals, theoretical
SEDs from Bruzual, to observed SEDs from known radio galaxies. 

The first use of the term in the title was by Loh and Spillar (1986):
{\it Photometric Redshifts of Galaxies}. This work was pioneering
in its use of CCDs to reach quite faint limits of $I \sim 21.5$, along
with the use of 6 medium-band filters and again $\chi^2$ template
fitting, but only to three observed {\it local} SED templates to
represent all galaxy types at all redshifts.

The use of multicolors to estimate redshifts actually predates the
above papers and the true literature on the subject is significantly
more extensive, but due to the ambiguity of the term ``photometric
redshifts'' as detailed in the next section, many workers in this
arena chose not to use this term.  Probably the first use of
multicolor photometry for redshift estimation was by Baum (1962). He
obtained photoelectric photometry through 9 medium-wide filters of
galaxies in cluster 3C395, assumed that the SEDs were that of
ellipticals, and obtained an estimate of redshift $z \sim 0.44$, quite
close to the eventual spectroscopic value of 0.46.

Nearly two decades passed before multicolor
photometry was used to estimate redshifts. Butchins (1981, 1983) used
UK Schmidt plate $BVR$ photometry that reached $B \sim 22$. Because of
the overlap in $BVR$ colors of low-redshift ($z \la 0.1$) early-type
galaxies with higher redshift ($z \sim 0.4$) later-type galaxies,
Butchins applied probablility constraints on luminosities that are
qualitatively similar to what has more recently been termed
``Bayesian''. I was able to reach fainter limits ($B \sim 24$) by
using KPNO 4-m plates and had both superior redshift accuracy $\delta
z \la 0.05$ and far less degeneracy by exploiting a set of filters
(roughly $UBRI$) with much longer wavelength coverage (Koo 1981, 1985,
1986). Since Baum is not here and Butchins is no longer in astronomy,
I qualify as the old-timer in the field at this workshop.

Another important set of papers that does {\it not} use the term
``photometric redshifts'', and yet constrains redshifts from broadband
photometry, relies on the so-called ``Lyman-break technique'' (see
contribution by Steidel). The basic idea is to discern very high
redshift galaxies via a significant drop in the bluest band for
galaxies with otherwise very blue colors in two or more redder
bands. This situation occurs when the Lyman break enters, e.g., the
$U$ band, at redshifts $z \sim 3$. Since the apparent drop in the $U$
band flux (or especially in redder bands at higher redshifts) is also
affected by the intergalactic Lyman-alpha forest depression, a more
accurate term should be ``Lyman-drop technique''. Early examples of
its use include papers by Partridge (1974), Koo(1986), Cowie (1988),
Majewski (1988, 1989), Guhathakurta et. al. (1990), and Steidel \&
Hamilton (1993).

\section{Working Definition of Photometric Redshifts}

I suggest the following: photometric redshifts are those derived from
{\it only} images or photometry with spectral resolution
$\lambda/\Delta \lambda \la 20$.  My choice of 20 is intended to
exclude redshifts derived from slit and slitless spectra, narrow band
images, ramped-filter images, Fabry-Perot images, Fourier transform
spectrometers, etc.

This definition still leaves a wide range of approaches to obtain
redshifts, examples of which include:

\begin{enumerate}

\item Spatial Correlations: Galaxies are assumed to have, {\it
statistically}, the redshifts of their neighbors in apparent close
pairs, groups, and especially the cores of rich clusters. Photometric
redshifts is an unconventional term to apply in these cases and should
thus probably be avoided.

\item Magnitudes: Although using just magnitudes alone to estimate
redshifts might be expected to work only for standard candles, such as
the brightest cluster galaxies in $R$ or strong-flux radio galaxies in
$K$, the convolution of accessible volumes with the shape of the
luminosity function of galaxies results in a fairly tight correlation
between magnitudes and redshift, even for more common field galaxies
(see the redshift-magnitude plot Fig. 4 of Koo \& Kron 1992).  For any
of the following techniques, the addition of magnitudes can thus be
expected to provide additional constraints on the redshift probability
distribution (see, e.g., Connolly et. al. 1995). Just as in the
previous approach, the use of the term ``photometric redshifts'' would
be quite unconventional, but technically accurate. 

\item One Color: In situations where the SED is known or unique, one
color may be sufficient to estimate a redshift. Practical applications
include redshift estimates for the reddest galaxies in clusters or
field and also very high redshift searches, especially when at least
one of the two bands is in the near-infrared. A single-color is
probably the minimal information to qualify the use of the term
``photometric redshift'' as understood by most astronomers today.

\item Two or More Colors: The use of three or more filters is
necessary to break the degeneracy between instrinsic color and
redshift. This approach is probably the most commonly accepted one
when referring to photometric redshifts that employ optical bands, and
works surprisingly well even with only three bands (Straizys \&
Sviderskiene 1983). This is because most galaxies (at least locally)
occupy only a small fraction of the possible multicolor volume (see
Fig. 2a); galaxy spectra are often composites of old and young stellar
populations which result in bowl-shaped or U-shaped SED (see Fig. 1
and note the more bowl-shaped value of the z = 0 locus compared to
that of its consituent stars); the pivot point of this curvature lies
near the 4000\AA \ break and is roughly independent of the galaxy's
average color \footnote{This coincidence of the 4000\AA \ break being
where galaxy SEDs have a pivot point results in the common
misconception that photometric redshifts work because of the 4000\AA \
break itself.}; and  bowl-shaped spectra, when redshifted,
result in moving the iso-z loci in a direction perpendicular to these
loci (see Fig. 1). Obviously one needs longer wavelength bands to
sample the pivot point near 4000\AA \ as one wishes to discern higher
redshifts (see Fig. 1). Nature has been kind: if galaxy spectra had
instead blackbody shapes of different temperatures or power-laws in
shape, we would not be able to separate redshifts using multicolors.

\item Surface Brightness: If the intrinsic surface
brightness is roughly constant or slowly varying with time, as might
be expected for large spirals that undergo largely constant star
formation rate histories, the $(1+z)^4$ surface brightness dimming can
be exploited to yield redshifts.

\item One Color with Light Profiles: In principle, the light
profile might yield the type of galaxy (e.g., $r^{1/4}$ profiles might imply a
luminous spheroidal  or bulge to disk ratios might
suggest a probable galaxy type), which in turn might be assumed to have a
unique SED so that a single color is sufficient for estimating a
redshift (Sarajedini et. al. 1999).

\item One Color with Image Structure: This is a generalization of the
last two approaches. Anderson et. al. (1996) has already been
exploring the correlations of galaxies among color, surface
brightness, and image concentration. This combination appears to be
useful to improve the efficiency of gathering photometric redshifts by
not requiring deep $U$ photometry, which is very costly in telescope
time.

\item etc.

\end{enumerate}

Whether the term ``photometric redshifts'' should include such a
diversity of approaches is debatable, but what should be obvious is
that astronomers have yet to exploit the wealth of additional
photometric/structural information in refining redshift estimates from
image data alone.

\section{Methodology}
\subsection{Template versus Empirical Fitting}

The most popular method for estimating redshifts from a set of
photometric measurements is the $\chi^2$ (or maximum likelihood)
template-fitting method adopted by Puschell et. al (1982), largely
because the method is simple, does not rely on having any
spectroscopic redshifts, and needs only a few templates (empirical or
theoretical) to yield results. With the availability of multicolor CCD
photometry, almost anyone can get into the photometric redshift
business today. The Achilles heel of the technique is of course the
template set. Empirical SEDs are relatively few, almost exclusively
confined to local galaxies whose SEDs may not reflect those of more
distant galaxies, and which are not necessarily representative in
luminosity, dust, inclination, morphology, etc. to the targets of
interest. While theoretical SEDs avoid some of these problems, they
may not be correct and do not yield information on the probability
distribution for the estimated redshifts, since this information
relies on having the probability distribution of the SEDs.

To show the level of uncertainty between empirical and theoretical
SEDs, see Fig. 1.  We note that the empirical spectra (which have
traditionally been based on only a handful of measured integrated
spectra) differ significantly (0.05 or more in z) from the
theoretical at all redshifts. 

\begin{figure}
\includegraphics{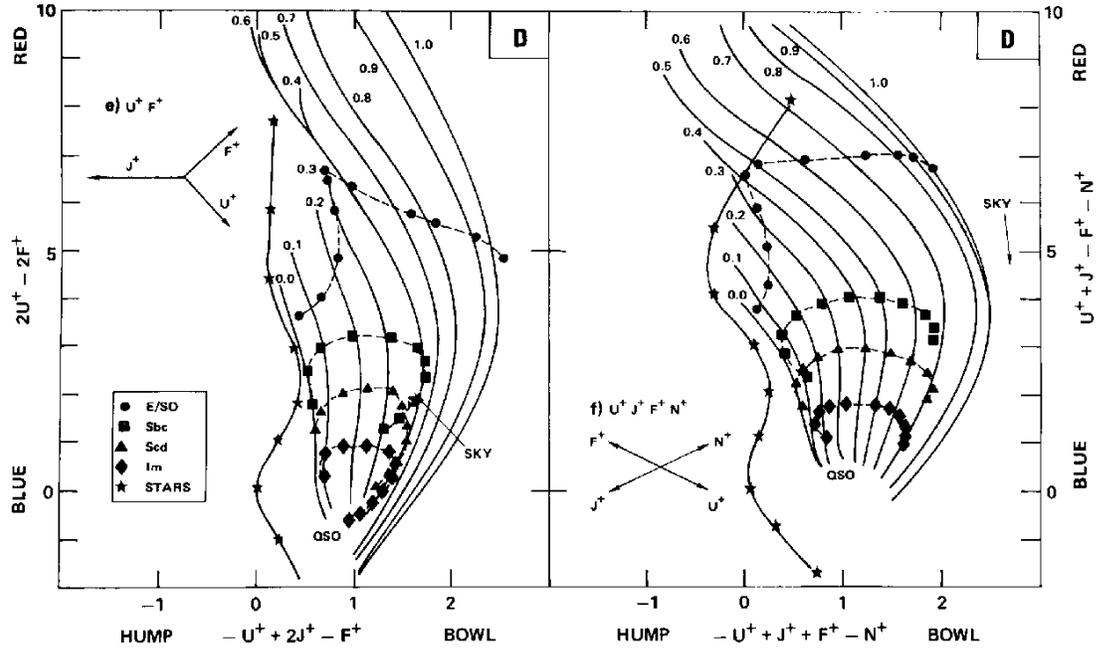}
\vspace{3.5in}
\caption{These multicolor plots show loci of constant redshift
from $z = 0.0$ to 1.0 in intervals of 0.1. The roughly vertical lines
are based on theoretical SEDs, while the roughly horizontal dashed
lines show the trajectory taken by empirical spectra of several types
of galaxies, with symbols spaced 0.1 in redshift. 
The left-hand figure is one where the normal two-color
plot (\ub \ vs $B-R$) is converted to one measuring the average color
($U^+-F^+$ equivalent to $U-R$) on the vertical axis and the curvature
of the spectrum ($-U^++2J^+-F^+$ equivalent to $-U+2B-R$) on the
horizontal axis, where humped and bowl shapes are indicated.  The
right-hand plot adds the $N^+$ band ($\sim I$), which extends the
wavelength range and thus improves both the color discrimination on
the vertical axis and the curvature or shape discrimination on the
horizontal.  The reader is referred to the caption of Fig. 1 in Koo
(1985) for more details. 
 }
\end{figure}

In contrast, the empirical fitting method (e.g., Connolly
et. al. 1995, Wang et. al. 1998) uses a large enough pool of
spectroscopic redshifts to calibrate the relationship of colors and
magnitudes to redshifts. By construction, this method should provide
accurate redshifts, but more importantly, it should also yield
realistic  estimates of the redshift
errors. To achieve this in practice, however, requires a sufficiently
complete set of known redshifts to the depth of the desired photometry. 

\subsection{Redshift Errors} 

Except for the mean and rms error when compared to spectroscopic
redshifts, photometric redshift errors remain poorly characterized.
In particular, the position, shape, and asymmetry of the redshift
error distribution should be derived; the sources of the errors and
especially whether they are due to random measurement errors,
intrinsic dispersions in the SEDs of galaxies, or unknown systematics
should be tracked down; and, ideally, the dependence of the errors on
a variety of other parameters (color, luminosity, morphology,
structure, redshift, environment, etc.)  should also be measured. More
importantly, these redshift uncertainties  should be incorporatedly explicitly
into the analysis, not by using the derived maximum likelihood or
minimum $\chi^2$ value of the photometric redshift for each galaxy,
but rather the full probability distribution. This idea has already
been incorporated for the C-method of deriving the luminosity function
from photometric redshifts (SubbaRao et. al. 1996), but this approach
should be adopted more universally by others in most statistical
analyses.

Of special concern in the area of redshift errors is the likelihood
both that the SED's of distant galaxies are different on average from
those today and that the intrinsic dispersion in the SEDs are greater.
If true, we would expect larger random and systematic errors for
galaxies at higher redshifts.  Figures 2 and 3 are telling and
sobering. More specifically, Figure 2b shows that morphologically
peculiar galaxies exhibit a greater spread of values {\it
perpendicular to the line} in the $UBV$ two-color plots than that of
more normal galaxies shown in Figure 2a (Larson and Tinsley
1978). Interactions and mergers appear to be a primary cause of these
peculiarities, so if indeed such galaxies are more common in the past,
we should expect greater photometric redshift errors at higher
redshifts that result from the greater spread in intrinsic SEDs
alone. Larson and Tinsley (1978) explain the wider spread as the
result of a greater diversity in the star formation histories of
galaxies, namely brief starbursts interrupting an otherwise more quiet
or more smoothly varying star formation history. Figure 3 shows the
possible tracks of galaxies with varying amounts of such bursts in the
$UBV$ two-color diagram; note in particular the regions {\it
perpendicular} to the wide, solid line on top of which most local,
morphologically-normal galaxies lie. Such deviations result in
systematic errors in photometric redshifts.

\begin{figure}
\plotone{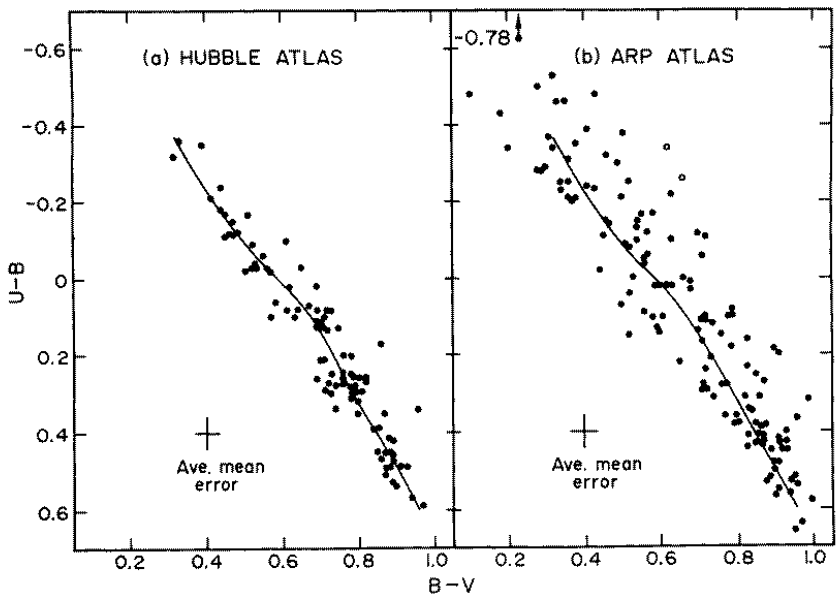}
\caption{
$UBV$ two-color plots showing the colors of (a) morphologically-normal
local galaxies versus those that are (b) peculiar 
(from Fig. 1 of Larson and Tinsley 1978, to which the interested reader is
referred for details of the sample selection).
}
\end{figure}

\begin{figure}
\plotone{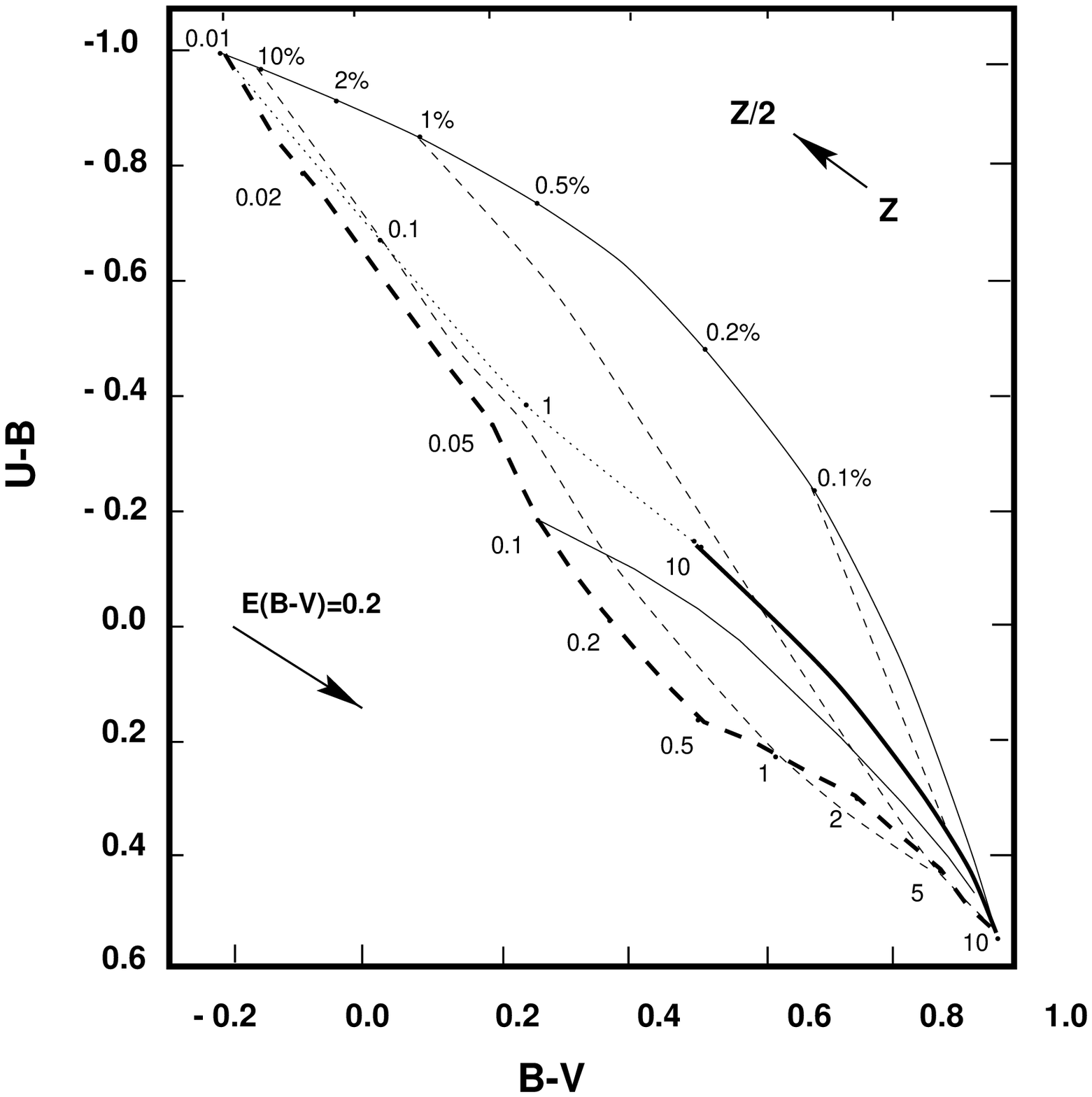}
\caption{
$UBV$ two-color plot showing the expected locations and trajectories
of galaxies with different star-formation histories (from Fig. 18 of
Tinsley 1980, to which the reader is referred for further
details). The thick solid line is where galaxies are predicted to lie
if they represent a range of smoothly decreasing rates of star
formation; as seen in Fig. 2, most local galaxies overlap this locus. 
The light dotted line is the track for constant star formation
over the indicated time (Gyr). The remaining tracks show brief starbursts
with different strengths (as \% of mass of a 10 Gyr old underlying stellar
population) and after indicated periods (Gyr).
}
\end{figure}

\subsection{Redshift Calibration Samples}

Clearly what is desired for significant progress in our use and
understanding of photometric redshifts is a large pool of
spectroscopic redshifts that span the full range of redshifts, depth,
photometric bands, etc. In 1985, we had about 100 spectroscopic
redshifts ($z \la 0.6$) to $R \sim 21$ from 4-5m class telescopes to
study photometric redshifts in four bands $UBRI$ (Koo 1985), while
roughly the same number has been measured with the Keck 10-m to study
the Hubble Deep Field to fainter limits $R \ga 23.5$ and a wider range
in redshifts ($z \sim 0.1 - 5.6$). Our current lack of good
calibration is, however, changing dramatically. At the lowest
redshifts of $z \la 0.2$, the Sloan Digital Sky Survey will yield
nearly $10^6$ spectra for a photometric sample that includes 5
filters. At intermediate redshifts between $z \sim 0.2$ to 0.7, CNOC2
will have 6000 spectroscopic redshifts for calibrating their 5 band system (Lin
et. al. 1999). Keck will be providing over 1000 redshifts for $z \sim
0.7$ to over $1$ and $\sim 2.5$ to 4 for passbands that include some $K$,
while handfuls of redshifts are coming in for the desert at $z \sim
1.3$ to over 2 and the highest redshifts $z \ga 5$. With these samples to
calibrate the photometric redshift system and especially its errors,
photometric redshifts will finally rest on a much more secure
foundation.

\section{Applications}

The scientific potential of using photometric redshifts has been
recognized for a long time, especially during an earlier era when
redshifts for large samples of galaxies fainter than $B \sim 21$ were
difficult or impractical, while multicolor photometry easily sampled
the distant universe. The following list is meant to be illustrative,
rather than representative or exhaustive, of pioneering projects that
were based on older multicolor photographic photometry or small-format
CCDs. The topics of other papers from this workshop provide a more
current view of how photometric redshifts are being applied in a new
era of HST images, large mosaic CCD cameras, and 8-10m telescopes.
The use of photometric redshifts has not only undergone a recent
revival, but it is also rapidly becoming a crucial tool of many
programs in mainstream observational cosmology.

\begin{itemize}

\item Searches for primeval galaxies at redshifts $z \ga 3$ via the
$U$ band detecting the Lyman break (Cowie 1988, Cowie \& Lilly 1989;
Guhathakurta et. al. 1990; Koo 1986; Majewski 1988, 1989; Steidel \&
Hamilton 1993) or even via data further to the red to probe higher
redshifts (Partridge 1974).

\item Searches for high redshift QSO's (see review by Warren \& Hewett
1990) or distant radio galaxies (Puschell et. al. 1982; van der Laan
1983).

\item Studies of the evolution of field galaxies (Butchins 1983; Cowie
et. al. 1988, 1990; Guiderdoni 1987; Koo 1986; Lilly et. al. 1991) or
their luminosity function (SubbaRao et. al. 1996)

\item Discrimination of cluster members and superclusters at moderate
redshifts (Connolly et. al. 1996; Koo 1981, 1986, Koo et. al. 1988)

\item Estimate of the geometry of the universe via the volume test
(Loh and Spillar 1986)

\end{itemize}

%

%
%


\acknowledgments

I would like to thank Ray Weymann for letting me be an old-timer,
albeit only briefly, and his conference team for organizing such a fun
workshop. With an old-timer brain, I have also probably missed a
number of important old references, and to their authors, my apologies
for any oversights. This work has been partially supported by NSF
AST-9529098 and by NASA through grant number AR-07532.01-96 from the 
Space Telescope Science Institute, which is operated by AURA, Inc.,
under NASA contract NAS 5-26555.

%

\end{document}